\documentclass[aps,amsmath,amssymb,11pt,superscriptaddress]{revtex4}
\usepackage{palatino}
\usepackage{graphicx}
\usepackage{amsthm}
\usepackage{bm}

\newtheorem{lemma}{Lemma}

\newtheorem{problem}{Problem}

\newcommand\ket[1]{\ensuremath{|#1\rangle}}

\begin{document}

\title{\Large\bf The LU-LC conjecture is false}

\author{Zhengfeng Ji}
\affiliation{State Key Laboratory of Computer Science,
Institute of Software, Chinese Academy of Sciences, P.O.Box 8718, Beijing 100080, China}
\author{Jianxin Chen}
\author{Zhaohui Wei}
\author{Mingsheng Ying}
\affiliation{
State Key Laboratory of Intelligent Technology and Systems,
Department of Computer Science and Technology, Tsinghua University, Beijing 100084, China}

\begin{abstract}
  The LU-LC conjecture is an important open problem concerning the
  structure of entanglement of states described in the stabilizer
  formalism. It states that two local unitary equivalent stabilizer
  states are also local Clifford equivalent. If this conjecture were
  true, the local equivalence of stabilizer states would be extremely
  easy to characterize. Unfortunately, however, based on the recent
  progress made by Gross and Van den Nest, we find that the conjecture
  is false.
\end{abstract}

\maketitle

\section{Introduction}
\label{sec:intro}

The stabilizer formalism is a group-theoretic framework originally
devised to systematically analyze various quantum error-correcting
codes~\cite{Got97} and has found applications in other areas of
quantum information processing, such as the one-way computation
model~\cite{RB01,DKP07} and the quantum sharing of classical
secrets~\cite{CL04}.

One of the key features of stabilizer states defined by the stabilizer
formalism is the presence of high-degree multipartite entanglement in
them. A natural approach to studying the properties of the
entanglement in stabilizer states is to investigate their local
equivalences. Mainly, three types of local equivalences have been
studied with respect to stochastic local operations and classical
communication (SLOCC), local unitary (LU), and local Clifford
operations (LC). In the following, we will call two states SLOCC (LU,
LC) equivalent if they can be transformed to each other by means of
SLOCC (LU, LC) operations.

Local Clifford operations are local unitary operations that map the
Pauli group to itself under conjugation. Therefore, the action of a
local Clifford operation on stabilizer state can be represented as a
simple way of translating the stabilizers. Because of this close
relationship between the stabilizer formalism and the Clifford group,
the LC equivalence of stabilizer states has been studied thoroughly in
the literature. For example, a polynomial time algorithm has been
found to decide whether two stabilizer states are LC
equivalent~\cite{NDM04a} and the action of local Clifford group on
graph states (an important subset of stabilizer states) has been
translated into elementary graph transformations characterized by a
single rule~\cite{NDM04b}. Specifically, the classification of LC
equivalent stabilizer states has been performed systematically up to
$12$ qubits~\cite{HEB04,Dan05,DP06}.

Stabilizer states have extremely symmetric structures which put strong
restrictions on the local unitary that can map one stabilizer state to
the other. In fact, it has been conjectured for several years that any
two LU equivalent stabilizer states must also be LC equivalent (the
LU-LC conjecture, listed as the 28th open problem in quantum
information~\cite{Sch05}). If this conjecture were true, all three
local equivalence of stabilizer states would be the same, as it has
already been proven that two stabilizer states are SLOCC equivalent if
and only if they are LU equivalent~\cite{NDM04}. Moreover, deciding
whether two states are locally equivalent would be efficient and the
local equivalence of stabilizer states could be described as purely
graph theoretic terms. The conjecture has been proved for large
subclasses of stabilizer states in Refs.~\cite{NDM05,ZCCC07} and is
further supported by the results obtained in Ref.~\cite{NDM05b}.

In this paper, however, we will present a counterexample to the LU-LC
conjecture which disproves the conjecture in general. Our result is
build upon the recent progress that transforms the conjecture to a
simpler problem~\cite{GN08}. In addition to our heuristics for the
construction of counterexamples, we also present some of the special
cases in which LU and LC equivalence are the same.

The organization of the this paper is as follows. In the next section,
we introduce the basics of stabilizer formalism and some notations and
results used in this paper. Sec.~\ref{sec:simp} develops new
representation of the problem in the first part and then considers
some special cases of the LU-LC conjecture. The procedure of
generating counterexamples and one explicit example are given in
Sec.~\ref{sec:counterexamples}. We conclude in
Sec.~\ref{sec:conclusion} and discuss some possible future work on
this problem.

\section{Notations}
\label{sec:note}

Stabilizer states are quantum states described by a set of commuting
operators. This idea of representing states with operators has been
proved to be extremely useful in the theory of quantum information.
Let $\{I,X,Y,Z\}$ be Pauli matrices,
\begin{equation*}
  X = \begin{bmatrix}0&\phantom{-}1\\1&\phantom{-}0\end{bmatrix},\quad
  Y = \begin{bmatrix}0&-i\\i&\phantom{-}0\end{bmatrix},\quad
  Z = \begin{bmatrix}1&\phantom{-}0\\0&-1\end{bmatrix},
\end{equation*}
$\mathcal{G}$ be the Pauli group generated by them, and $I_n$ be
$I^{\otimes n}$. Mathematically, a stabilizer of $n$ qubits is an
Abelian subgroup of $\mathcal{G}^{\otimes n}$ that does not include
$-I_n$. When the subgroup has cardinality exactly $2^n$, there will be
a unique quantum state determined by it as the simultaneous fix point
of all the operators in the subgroup. For example, the EPR pair
$(\ket{00}+\ket{11})/\sqrt{2}$ is stabilized by the group generated by
$\{X\otimes X, Z\otimes Z\}$. For any graph $G$, the
corresponding graph state is a special stabilizer state with
\begin{equation*}
  X_v \bigotimes_{u\in N(v)} Z_u,\quad v\in V(G)
\end{equation*}
as its stabilizer, where $N(v)$ is the neighbor set of vertex $v$. It
is known that any stabilizer state is LC equivalent to some graph
state~\cite{NDM04b}. Therefore, the LU-LC conjecture for stabilizer
states and graph states are the same problem. For more details on the
power of stabilizer formalism and graph states, the readers are
referred to Refs.~\cite{Got97,HDE+06}.

There are special structures in the amplitudes of stabilizer states
expanded in the computational basis. It is proved that any
stabilizer state can be written as
\begin{equation}
  \label{eq:expansion}
  \frac{1}{\sqrt{|T|}} \sum_{x\in T} i^{l(x)} (-1)^{q(x)} \ket{x},
\end{equation}
where $T$ is an affine space of $\mathbb{F}^n_2$, $l(x)$ is linear in
$x_j$ with addition modulo $2$, and $q(x)$ is a quadratic function of
$x_j$'s. Conversely, any state having the above form is a stabilizer
state~\cite{DM03}.

The main topic in this paper is to study local equivalences of
stabilizer states, especially the LU equivalence and LC equivalence.
Clifford operators are $2$ by $2$ unitary operators $U$ such that
$U\mathcal{G} U^\dagger = \mathcal{G}$. Up to a global phase, the
Clifford operators form a finite subgroup of $U(2)$. We say two
stabilizers $\ket{\psi_0}$ and $\ket{\psi_1}$ are LC equivalent if
there exists Clifford operators $U_j$ such that
\begin{equation*}
  \bigotimes_{j=1}^n U_j \ket{\psi_0} = \ket{\psi_1}.
\end{equation*}

In the following, we define several problems, each of which is given
an abbreviated name for later references. The first one is the
original LU-LC conjecture itself, then a restricted case of it,
DLU-LC. The third is the DLU-DLC problem and the last is quadratic
form phase problem (QFP). One can see the close relationship between
QFP and the other three problems from Eq.~\eqref{eq:expansion}. The
previously known relation of these problem is as follows. DLU-LC is
shown to be equivalent to the LU-LC conjecture~\cite{GN08,ZCC07}.
DLU-DLC and QFP are equivalent and they imply the LU-LC conjecture
according to Ref.~\cite{GN08}. All of these four statements are false
as indicated by our counterexample given in
Sec.~\ref{sec:counterexamples}.

\begin{problem}[LU-LC]
  Every two LU equivalent stabilizer states are also LC equivalent.
\end{problem}

\begin{problem}[DLU-LC]
  Every two stabilizer states that can be mapped onto each other by
  means of a diagonal local unitary, are LC equivalent.
\end{problem}

\begin{problem}[DLU-DLC]
  If two stabilizer states can be mapped onto each other by means of a
  diagonal local unitary, then also by a diagonal local Clifford
  operation.
\end{problem}

\begin{problem}[QFP~\cite{GN08}]
  Let $S$ be a linear subspace of $\mathbb{F}_2^n$, and
  $Q:\mathbb{F}_2^n\rightarrow \mathbb{F}_2$ be a quadratic function.
  If there exists complex phases $\{c_i\}$ such that
  \begin{equation}
    \label{eq:qf}
    (-1)^{Q(x)} = \prod_j^n c_j^{x_j},\qquad \text{for every }x\in S,
  \end{equation}
then the phases can be chosen from $\{\pm 1, \pm i\}$.
\end{problem}

\section{Simplification of the QFP Problem}
\label{sec:simp}

Let us start by simplifying the QFP problem. We will give a linearized
version of QFP in the following. It's easy to obtain linear
representations of QFP by, for example, enumerating all possible $x\in
S$ and resulting in a system of linear congruence equations. But this
method does not give us much information on how to tackle the problem.
What we present in the following is a more symmetric linear
representation that leads to the construction of counterexamples of
QFP.

\subsection{Symmetric Linear Representation}
\label{sec:linear}

There are two main mathematical objects in the problem, the quadratic
function $Q(x)$ and the subspace $S$. We will find a representation of
$Q(x)$ first and then analyze the structure of $S$. Based on this
analysis, some special cases of QFP will be proved. We will also
obtain an approximate restatement of the general QFP problem that
leads to the construction of counterexamples.

First, noticing that the linear terms in $Q(x)$ can be moved to the
right hand side of Eq.~\eqref{eq:qf} without changing the problem, we
can consider quadratic forms only. There is a trivial one-to-one
correspondence between quadratic forms over $\mathbb{F}_2$ and simple
graphs. For each $Q(x)$, one can identify it with a graph
$\mathcal{Q}$ with vertex set $V=\{1,2,\ldots,n\}$ and edge set
\begin{equation*}
  E(\mathcal{Q}) = \{(i,j)\mid x_i x_j\text{ is a term in } Q(x)\}.
\end{equation*}
Each subset $A$ of $V$ defines a subgraph
$\mathcal{Q}\vert_{\displaystyle A}$ whose vertex set and edge set are
$A$ and $E(\mathcal{Q}) \cap A^2$ respectively. With the above
definitions, one can verify that
\begin{equation}
  \label{eq:subgraph}
  Q(x) =
  \left| E(\mathcal{Q}\vert_{\displaystyle I_x}) \right|,
\end{equation}
where $I_x$ is the indicator set $\{j\mid x_j=1\}$.

Next, to analyze the structure of $S$, we choose a basis
$\{\xi^1,\xi^2,\ldots,\xi^d\}$ of $S$. For each $x\in S$, we can
represent it as the linear combination
\begin{equation*}
  x = \sum_{k=1}^d h_k \xi^k.
\end{equation*}
Let $h\in \mathbb{F}^d_2$ be the vector whose $k$-th coordinate is
$h_k$ and we label the above $x$ with it as $x^h$. The size of $S$ is
therefore $D=2^d$.

\begin{figure}[!hbt]
  \centering
  \includegraphics{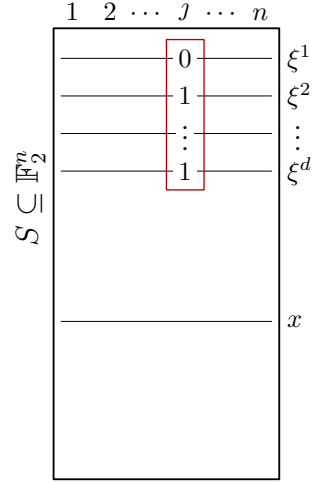}
  \caption{Illustration of patterns}
  \label{fig:pattern}
\end{figure}

We introduce a new concept called {\it pattern} of a position in $S$.
The pattern for position $j$ is defined to be the vector $m\in
\mathbb{F}^d_2$ whose $k$-th coordinate $m_k$ equals to $\xi^k_j$. See
Fig.~\ref{fig:pattern} for a illustration of this definition. Patterns
are basis dependant but the choice of basis does not change the
following analysis.

Let $A_m$ be the set of positions having pattern $m$, that is,
\begin{equation*}
  A_m = \{ j\mid m_k = \xi^k_j \text{ for all } k\}.
\end{equation*}
The collection $\{A_m\}$ is obviously a partition of the vertex set
$V$ of graph $\mathcal{Q}$. For any $j\in A_m$, one can calculate the
$j$-th coordinate of $x^h$ as
\begin{equation}
  x^h_j = \sum_{k=1}^d h_k \xi^k_j =
  \sum_{k=1}^d h_k m_k \stackrel{\textrm{def}}{=} \langle h, m\rangle,
\end{equation}
where the summation is taken over $\mathbb{F}_2$, or modulo $2$. That
is, the value of $x^h$ at position $j$ is determined solely by the
pattern of the position. We can define variable $x_{[m]}$ to be a
representative variable of those having pattern $m$.

Consider a special case of QFP where $A_m$ is nonempty for all
patterns $m\in\mathbb{F}^d_2-\{0\}$. We will prove that in this case
QFP is true and the details of the calculation will be useful even
when we deal with the general QFP problem.

As all $A_m$'s are nonempty, we can rewrite Eq.~\eqref{eq:qf} in the
QFP problem by replacing the variables with their representatives in
the right hand side,
\begin{equation*}
  (-1)^{Q(x)} = \prod_{m\ne 0} C_{[m]}^{\displaystyle x_{[m]}},
  \qquad \text{for every }x\in S,
\end{equation*}
where $C_{[m]}$ is the product of all $c_j$ for $j\in A_m$. For a
specific $x^h\in S$, this means that
\begin{equation}
  \label{eq:qf-pat}
  (-1)^{Q(x^h)} = \prod_{m\ne 0} C_{[m]}^{\langle h, m\rangle}.
\end{equation}
From Eq.~\eqref{eq:subgraph} and the fact that
\begin{equation}
  I_{x^h} = \bigcup_{m:\langle m, h\rangle=1} A_m,
\end{equation}
we have
\begin{equation*}
  \begin{split}
    Q(x^h) & = \sum_{m:\langle m, h\rangle=1}
      \left| E(\mathcal{Q}\vert_{\displaystyle A_m}) \right| +
      \sum_{m<m':\langle m, h\rangle=1,\langle m', h\rangle=1} E_{mm'}\\
           & = \sum_m \langle m, h\rangle
      \left| E(\mathcal{Q}\vert_{\displaystyle A_m}) \right| +
      \sum_{m<m'} \langle m, h\rangle \langle m', h\rangle E_{mm'},
  \end{split}
\end{equation*}
where $E_{mm'}$ counts the number of edges between vertex sets $A_m$
and $A_{m'}$ in graph $\mathcal{Q}$. The first part of the above
summation can be omitted as it can be absorbed into the right hand
side of Eq.~\eqref{eq:qf-pat} by changing the values of $C_{[m]}$
appropriately. Equation~\eqref{eq:qf-pat} is therefore reduced into
the following form,
\begin{equation}
  \label{eq:qf-simp}
  (-1)\,\,\,^{\displaystyle\sum\limits_{m<m'} \langle m, h\rangle \langle m', h\rangle E_{mm'}} = \prod_{m\ne 0} C_{[m]}^{\langle m, h\rangle}.
\end{equation}

Let $r_{[m]}$ be the real number satisfying
\begin{equation}
  C_{[m]} = i^{r_{[m]}}.
\end{equation}
By taking logarithm on both side of Eq.~\eqref{eq:qf-simp}, one gets
\begin{equation*}
  \sum_{m\ne 0} \langle m, h\rangle r_{[m]} \equiv
  2 \sum\limits_{m<m'} \langle m,h\rangle\langle m',h\rangle E_{mm'}\pmod{4}.
\end{equation*}
This equation holds for all $h\in\mathbb{F}^d_2$, ant it actually
specifies a system of $D-1$ equations where $D=2^d$. By defining $D-1$
by $D-1$ matrix $G = (\langle i,j\rangle)$ and matrix $T$ with element
\begin{equation}
  \label{eq:T}
    T_{i,(j,k)} = \langle i,j\rangle \langle i,k\rangle,\quad
    i,j,k\in\mathbb{F}^d_2-\{0\},j<k,
\end{equation}
we can write the system of equations succinctly as
\begin{equation}
  \label{eq:qf-linear}
  G\vec{r} \equiv 2 T\vec{e} \pmod{4}
\end{equation}
where $\vec{r}$ and $\vec{e}$ are vectors consisting of $r_{[m]}$ and
$E_{mm'}$ respectively.

Next, we calculate $G^{-1}$ and the product $2G^{-1}T$ based on the
following lemma.
\begin{lemma}\label{lemma:identities}
  The following identities hold with the indexes traversing
  $\mathbb{F}^d_2-\{0\}$.

  \begin{equation}
    \sum_j \langle i,j \rangle = D/2,
  \end{equation}
  \begin{equation}
    \sum_j \langle i,j \rangle \langle j,k \rangle =
    \begin{cases}
      \ D/2\quad & i=k\\
      \ D/4\quad & \text{otherwise}
    \end{cases}
  \end{equation}

  For $k\ne l$, we have
  \begin{equation}
    \sum_j \langle i,j \rangle \langle j,k\rangle
    \langle j,l\rangle =
    \begin{cases}
      \ D/4\quad & i=k\text{ or }i=l\\
      \ 0\quad & i=k\oplus l\\
      \ D/8\quad & \text{otherwise}
    \end{cases}
  \end{equation}
\end{lemma}

We prove the last identity only, others are simpler and can be dealt
with similarly. When $i=k$ or $i=l$ the result follows from the second
identity. In the case of $i=k\oplus l$, there does not exist $j$ so
that $\langle j,i\rangle$, $\langle j,k\rangle$, $\langle j,l\rangle$
are all $1$. What remains to show is when $i\ne k$, $i\ne l$, $k\ne l$
and $i\ne k\oplus l$. Consider a $3$ by $d$ matrix $F$ with $i,k,l$ as
its rows, then it is a matrix of rank $3$ under the above conditions.
$j$ contributes $1$ to the sum when it is a solution of
$Fj=(1,1,1)^T$. The number of solutions is obviously $D/8$.

With the above identities, it's easy to check that $G^{-1}$ is given by
\begin{equation}
  \label{eq:ginverse}
  G^{-1} = \frac{2}{D} (2\langle i,j\rangle-1),
\end{equation}
which is obtained by replacing the zero elements in $G$ with $-1$ and
multiplying a normalization constant. More importantly, the product
$2G^{-1}T$ is still an integral matrix with $1$ at $(i,(j,k))$ where
$i=j$ or $i=k$, $-1$ where $i=j\oplus k$ and $0$'s elsewhere. This
immediately proves the QFP problem in the special case where $A_m$'s
are all nonempty, as we can always choose $\vec{r}$ to be $2G^{-1}T
\vec{e}$, an integral vector. The proof is completed without even
referring to the condition that there exists a real solution of
$\vec{r}$ for Eq.~\eqref{eq:qf-linear} as promised by the QFP problem.

However, the general case QFP problem is much more complicated and, in
fact, false sometimes. When $A_m$ is empty for some pattern $m$, we
cannot introduce representative variable $x_{[m]}$ as there is no
corresponding variables. But in order to make use of the above
formalism, we keep the variables $x_{[m]}$ for all those missing
patterns $m$ and, at the same time, fix the corresponding $C_{[m]}$'s
to be $1$, or fix $r_{[m]}$'s to be $0$ equivalently. Hence, the QFP
problem is approximately reduced to the following statement in terms
of linear congruence equations (LCE). We have intentionally simplified
the problem by omitting the indication of places forced to be $0$ in
$\vec{r}$. This gives us a stronger statement than QFP, that is, the
LCE problem implies QFP and any counterexample of QFP also invalidates
LCE. Compared to the trivial linearization of QFP, we have actually
used more variables to make the representation more symmetric.

\begin{problem}[LCE]
  If the following equation (Eq.~\eqref{eq:qf-linear}) of $\vec r$ has
  a real solution,
  \begin{equation*}
    G \vec{r} \equiv 2 T \vec{e} \pmod{4}
  \end{equation*}
  then it also has an integral solution that preserves all zero
  entries in the real solution.
\end{problem}

\subsection{The Low-Rank Case}
\label{sec:lowrank}

In this subsection, we discuss the QFP problem with low-rank subspace
$S$. Namely, we want to show that for $d\le 5$, there is no
counterexamples of QFP. As QFP is a sufficient condition for LU-LC, it
follows that LU-LC conjecture holds for $d\le 5$.

We need only to prove that there is no counterexamples for LCE in this
case. First, any solution $\vec{r}$ of Eq.~\eqref{eq:qf-linear} can be
written as
\begin{equation*}
  \vec{r} = 2G^{-1}T\vec{e} + 4G^{-1}\vec{s},
\end{equation*}
for some integral vector $\vec{s}$, and vice versa. Substitute
Eq.~\eqref{eq:ginverse} into the second part,
\begin{equation*}
  \vec{r} = 2G^{-1}T\vec{e} + \frac{16}{D}G\vec{s} - \frac{8\sigma}{D}\vec{1},
\end{equation*}
where $\vec{1}$ is the vector whose entries are all $1$ and $\sigma$
is the summation of all entries in vector $\vec{s}$. When $d\le 3$ all
parts in the summation are integral. If $d=4$, only the last term can
be half integral. However, when this happens, no entry in $\vec{r}$ is
integral, and the LCE problem holds by choosing $\vec{s}=0$.

Consider now the case of $d=5$. Let $[x]$ be the largest integer not
exceeding $x$, and $\{x\}$ be $x-[x]$. We claim that if $\vec{r}$ is a
solution of Eq.~\eqref{eq:qf-linear}, the truncation $[\vec{r}]$ is
also a valid solution. It suffices to prove that, for any integral
$\vec{s}$,
\begin{equation}
  \label{eq:rounding}
  G \left[4 G^{-1} \vec{s} \right] \equiv 0\pmod{4}.
\end{equation}
Employing Eq.~\eqref{eq:ginverse} and a detailed discussion on the
four possible cases of $\sigma\pmod{4}$, we can reduce the above
equation to
\begin{equation*}
  G \left\{ \frac{1}{2}G\vec{s} \right\} \equiv 0\pmod{4},
\end{equation*}
which in turn follows easily from the second identity of
Lemma~\ref{lemma:identities} by noticing that
\begin{equation*}
  2 \left\{ \frac{1}{2}G\vec{s} \right\}
\end{equation*}
is a column of $G$.

We have shown that it is impossible to construct counterexamples of
the QFP problem when the rank of subspace $S$ is less than or equal to
$5$. Naturally, the next step is to investigate the case of $d=6$,
which will lead to counterexamples.

\section{Random generation of counterexamples}
\label{sec:counterexamples}

Notice that the ``zero-entry preserving'' requirement is essential in
LCE. Without this additional requirement, LCE will be true. Therefore,
we try greedily to find a real solution of $\vec{r}$ with as many zero
entries as possible. As the number of nonzero entries will be the size
of our counterexample, this approach also tries to minimize the size
of counterexamples.

Recall the key equation of the problem
\begin{equation*}
  \vec{r} = 2G^{-1}T\vec{e} + 4G^{-1}\vec{s}.
\end{equation*}
As the first part of the right hand side is integral, we will at least
need to find an integral vector $\vec{s}$ so that $4G^{-1}\vec{s}$ has
many integral entries in order to have a $\vec{r}$ with many $0$'s.
Conversely, once $4G^{-1}\vec{s}$ is chosen, we can try to cancel most
of its integral entries by choosing $\vec{e}$. And after $\vec{s}$ and
$\vec{e}$ are fixed, we can calculate $\vec{r}$ and verify whether
there is no integral solution that preserves zero entries in it. It is
a backward approach that chooses in order, $\vec{s}$, $\vec{e}$, and
$\vec{r}$. More details on this approach are given in the following.

We first choose $\vec{s_0}$ at random. But it generally does not give us
many integral entries in $4G^{-1}\vec{s_0}$. We can employ the rounding
technique once again previously used in the case of $d=5$. For $d=6$,
the following equation similar to Eq.~\eqref{eq:rounding} can be
proved
\begin{equation*}
  G \left[8 G^{-1} \vec{s_0} \right] \equiv 0\pmod{8}.
\end{equation*}
Therefore, there exists an integral $\vec{s_1}$ such that
\begin{equation*}
  \frac{1}{2}\left[8 G^{-1} \vec{s_0} \right] \equiv 4 G^{-1} \vec{s_1} \pmod{4}.
\end{equation*}
There will be generally more integers in $4G^{-1}\vec{s_1}$ than in
$4G^{-1}\vec{s_0}$ and we can use $\vec{s_1}$ instead of $\vec{s_0}$.

In the next step, vector $\vec{e}$ is chosen to cancel the integral
entries in $4G^{-1}\vec{s_1}$. Thanks to the simple form of
$2G^{-1}T$, it is flexible to find $\vec{e}$ that fulfills the
requirement and, in addition, satisfies that $E_{mm'}=0$ when pattern
$m$ corresponds to an integral entry in $4G^{-1}\vec{s_1}$.

Vector $\vec{e}$ gives most of the useful information on the quadratic
form $Q(x)$, and we construct subspace $S$ by deleting columns of $G$
corresponding to zero entry positions in $\vec{r}$. This is how a
candidate counterexample is generated.

Finally, we need to verify whether the $Q(x)$ and $S$ obtained here
constitute a valid counterexample of QFP, that is, to verify that
there do not exist $c_j\in\{\pm 1,\pm i\}$ satisfying
Eq.~\eqref{eq:qf}. This can be done by a modified Gaussian elimination
method described below.

Let $A$ be the matrices contains all vectors in $S$ as its rows. The
problem of whether $c_i$ can all be chosen from $\{\pm 1, \pm i\}$ is
equivalent to whether an equation of the following form has an
integral solution
\begin{equation}
  \label{eq:axb}
  A x \equiv b \pmod{4}.
\end{equation}

The main difficulty in performing Gaussian elimination is that in
congruence equations we cannot perform division to normalize the
pivot. Fortunately, for equations of modular $4$, division operations
can be avoided. When performing the elimination procedure, try first
to find a row with odd pivot and if the pivot is congruent to $3$,
multiply $3$ on it. Otherwise, if we cannot find a row having odd
pivot, the division is also unnecessary as one can do elimination
simply by subtraction. Using this modified elimination method, we can
derive from Eq.~\eqref{eq:axb} a contradicting equation as
\begin{equation}
  0 \equiv 2 \pmod{4},
\end{equation}
under the assumption of the existence of an integral solution.

The above procedure to randomly generate different counterexamples is
implemented and is available at
http:/\!\!/\!arxiv.org/\!e-print/\!0709.1266 as a gzipped tar
(.tar.gz) file. Here is one of the counterexamples it generates.

In the example, $n=27$ and the subspace $S$ of $\mathbb{F}_2^{27}$ is
given by a set of basis as
\begin{equation}
  \label{eq:ce_s}
  \begin{split}
    \xi^1 & = 100010001010101010100011110\\
    \xi^2 & = 101010111001100000001010101\\
    \xi^3 & = 011001100111100111100110011\\
    \xi^4 & = 000111100000011001100001111\\
    \xi^5 & = 000000011111111000011111111\\
    \xi^6 & = 000000000000000111111111111.
  \end{split}
\end{equation}
One can see that every two columns (patterns) are different and the
columns are organized in an increasing order. $S$ has a rank of $6$
and consists $64$ elements.

The quadratic form $Q(x)$ is the summation of $11$ terms:
\begin{equation}
  \label{eq:ce_q}
    x_{1}x_{2} + x_{1}x_{3} + x_{1}x_{8} + x_{2}x_{4} +
    x_{2}x_{8} + x_{2}x_{16} + x_{3}x_{4} + x_{3}x_{8} +
    x_{3}x_{16} + x_{4}x_{8} + x_{8}x_{16}.
\end{equation}
We note that this is not directly generated by the random procedure.
In fact, we have find a $Q$ with smaller number of terms that has the
same value $Q(x)$ as the generated quadratic form for all $x\in S$.
Originally, the quadratic form contains more than $60$ terms. This
simplification is suggested by one of the referees of QIP 2008.

The QFP problem defined by the above $Q$ and $S$ has a solution where
the phases are powers of $e^{\pi i/4}$, but not in $\{\pm 1,\pm i\}$.
We give the exponents sequentially:
\begin{equation}
  \label{eq:ce_c}
    e = [3, 5, 7, 5, 1, 3, 5, 7, 1, 3, 5, 5, 3, 7, 3, 3, 7,
    1, 7, 3, 1, 5, 5, 5, 3, 5, 3].
\end{equation}

Our final aim is to disprove the LU-LC conjecture and what we have
already shown is that QFP is false generally. As proved in
Ref.~\cite{GN08}, QFP implies LU-LC. Therefore, QFP is false does not
logically guarantee that LU-LC is false. Fortunately, however, the
above counterexample of QFP can be transformed to a counterexample of
LU-LC. Although not proved, this transformation seems to work all the
time.

Here is the way we perform the transformation. Define two states
\begin{equation}
  \begin{split}
    \ket{S} & = \sum_{x\in S} \ket{x}\\
    \ket{Q,S} & = \sum_{x\in S} (-1)^{Q(x)}\ket{x}.
  \end{split}
\end{equation}
From Eq.~\eqref{eq:expansion}, we know that they are stabilizer
states. The corresponding QFP problem of the same $Q$ and $S$
indicates that they are LU equivalent, in fact, even DLU equivalent.
We need to show that they are not LC equivalent. This can be done
efficiently by using the LC equivalence decision
algorithm~\cite{NDM04a} for graph states after efficiently finding
graph states $\ket{G_S}$ and $\ket{G_{Q,S}}$ which are LC equivalent
to $\ket{S}$ and $\ket{Q,S}$ respectively. If the algorithm tells that
$\ket{G_S}$ and $\ket{G_{Q,S}}$ are not LC equivalent, nor are the
states $\ket{S}$ and $\ket{Q,S}$ (see Fig.~\ref{fig:trans}). The four
states in the diagram are in the same equivalence class under LU
criterion. But when LC equivalence are considered, the upper two and
the lower two belong to different classes.

\begin{figure}[!hbt]
  \centering
  \begin{tabular}[c]{rcl}
    $\displaystyle \ket{Q,S} = \frac{1}{\sqrt{|S|}}\sum_{x\in S}
    (-1)^{Q(x)} \ket{x}$ &
    $\stackrel{\text{\ \ LC\ \ }}{\longleftrightarrow}$ &
    $\ket{G_{Q,S}}$\\[1em]
    \rotatebox{90}{$\longleftrightarrow$}
    \raisebox{.5em}{$\stackrel{\text{LU}}{}$} & &
    \rotatebox{90}{$\longleftrightarrow$} 
    \raisebox{.5em}{$\stackrel{\text{{\bf NOT} LC}}{}$}\\
    $\displaystyle \ket{S} = \frac{1}{\sqrt{|S|}}\sum_{x\in S} \ket{x}$ &
    $\stackrel{\text{\ \ LC\ \ }}{\longleftrightarrow}$ &
    $\ket{G_{S}}$
  \end{tabular}
  \caption{State Transformation}\label{fig:trans}
\end{figure}

A better solution to show the non-LC equivalence of $\ket{Q,S}$ and
$\ket{S}$ would be a proof that DLU and LC imply DLC. This is open
generally, but it is pointed out by B. Zeng that when local unitaries
are all non-Clifford, just as in the above case, a proof can be found
using results on minimal support of stabilizers developed in
Ref.~\cite{NDM05}.

\begin{figure}[!hbt]
  \centering
  \includegraphics[width=.45\textwidth]{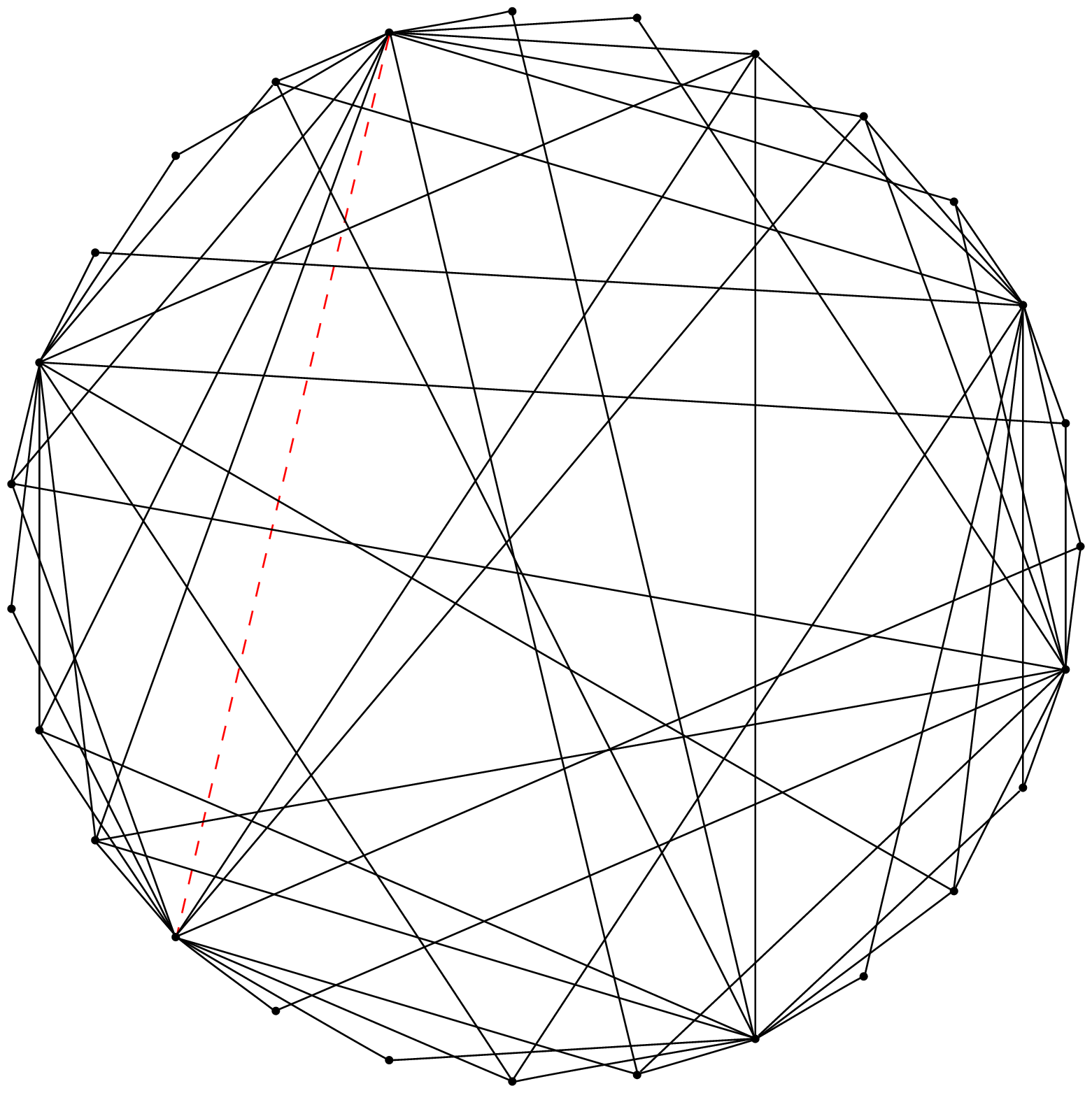}
  \caption{Corresponding Graph States}
  \label{fig:graphs}
\end{figure}

The two graph states $\ket{G_S}$ and $\ket{G_{Q,S}}$ have
corresponding graphs shown in Fig.~\ref{fig:graphs}. The graphs with
and without the dotted edge are graph $G_{Q,S}$ and $G_{S}$
respectively. Interestingly, two LU equivalent but not LC equivalent
graph states can differ only in on edge. Note that the simplification
of $Q(x)$ discussed above does not simplify or change the form of the
four stabilizer states in Fig.~\ref{fig:trans}.

\section{conclusion}
\label{sec:conclusion}

In summary, we have shown that the LU-LC conjecture is false by giving
explicit counterexamples of it. It is also clear that in order to
disprove the conjecture, we have to consider stabilizer states with
the rank of the support no less than $6$. This result leads us to
rethink about the local equivalence problem of stabilizers as most of
the previous work focus on proving the conjecture to be true.

The random procedure generates counterexamples of size $27$ and $35$.
Though not proved, we believe that $27$ is the smallest possible size
of counterexamples of LU-LC. Although the LU-LC conjecture is now
disproved, we still know little about the relation of LU and LC
equivalence and do not have an explicit understanding of why or when
LU equivalence differs from LC equivalence. When $d=6$, the random
generation procedure can find a counterexample in seconds, but it is
in fact not an efficient algorithm when $d$ is large or even when
$d=7$. That fact is we have never obtained a valid counterexample of
$d=7$ using the random procedure. Fortunately, larger scale
counterexamples, including those of $d=7$, have been found~\cite{GZ08}
motivated by the randomly generated counterexamples.

As LU and LC equivalences are now known to be different. It is also
challenging to ask whether there is an efficient algorithm deciding LU
equivalence for stabilizers or whether there is a graph theoretical
interpretation of LU equivalent graph states. This seems to be
difficult as local unitary operations are much less linked up with the
stabilizer formalism.

\section*{Acknowledgement}

We are thankful to David Gross, Maarten van den Nest, Bei Zeng, Markus
Grassl and the colleagues in the Quantum Computation and Information
Research Group of Tsinghua for helpful discussions. We note that a gap
in an earlier version of this paper was found by Andre Ahlbrecht and
communicated to the authors by David Gross. This work was partly
supported by the National Natural Science Foundation of China (Grant
Nos.~60721061, 60503001 and 60621062).

\bibliographystyle{abbrv}
\bibliography{LU-LC}

\end{document}